\documentclass[journal]{IEEEtran}
\usepackage{cite}
\usepackage{graphicx}
\usepackage{amsmath,amssymb,amsfonts}
\usepackage{algorithm, algorithmic}
\usepackage{subfigure}
\usepackage{multirow}
\usepackage{array}
\usepackage{color}
\usepackage{verbatim}
\usepackage{CJKutf8}

\begin{document}

\title{Deep Joint Source-Channel Coding for CSI Feedback: An End-to-End Approach}
%
%
%

\author{Jialong Xu,~\IEEEmembership{Student Member,~IEEE,}
        Bo Ai,~\IEEEmembership{Fellow,~IEEE,}
        Ning Wang,~\IEEEmembership{Member,~IEEE,}
        Wei Chen,~\IEEEmembership{Senior Member,~IEEE}
\thanks{\textit{(corresponding authors: Bo Ai; Wei Chen)}}
\thanks{Jialong Xu is with the State Key Laboratory of Rail Traffic Control and Safety, Beijing Jiaotong University, Beijing 100044, China (e-mail: jialongxu@bjtu.edu.cn).}
\thanks{Bo Ai is with the State Key Laboratory of Rail Traffic Control and Safety, the School of Electronic and Information Engineering, and the Frontiers Science Center for Smart High-Speed Railway System, Beijing Jiaotong University, Beijing 100044, China, also with the Research Center of Networks and Communications, Peng Cheng Laboratory, Shenzhen 518055, China, and also with the Henan Joint International Research Laboratory of Intelligent Networking and Data Analysis, Zhengzhou University, Zhengzhou 450001, China (e-mail: boai@bjtu.edu.cn).}
\thanks{Ning Wang is with School of Information Engineering, Zhengzhou University, Zhengzhou 450001,China (e-mail: ienwang@zzu.edu.cn).}
\thanks{Wei Chen is with the State Key Laboratory of Rail Traffic Control and Safety, Beijing Jiaotong University, Beijing 100044, China (e-mail: weich@bjtu.edu.cn).}

}

\markboth{Journal of \LaTeX\ Class Files,~Vol.~14, No.~8, August~2021}%
{Xu \MakeLowercase{\textit{et al.}}: Image Encryption for Deep Source Channel Coding}

\maketitle

\begin{abstract}
The increased throughput brought by MIMO technology relies on the knowledge of channel state information (CSI) acquired in the base station (BS). To make the CSI feedback overhead affordable for the evolution of MIMO technology (e.g., massive MIMO and ultra-massive MIMO), deep learning (DL) is introduced to deal with the CSI compression task. Based on the separation principle in existing communication systems, DL based CSI compression is used as source coding. However, this separate source-channel coding (SSCC) scheme is inferior to the joint source-channel coding (JSCC) scheme in the finite blocklength regime. In this paper, we propose a deep joint source-channel coding (DJSCC) based framework for the CSI feedback task. In particular, the proposed method can simultaneously learn from the CSI source and the wireless channel. Instead of truncating CSI via Fourier transform in the delay domain in existing methods, we apply non-linear transform networks to compress the CSI. Furthermore, we adopt an SNR adaption mechanism to deal with the wireless channel variations. The extensive experiments demonstrate the validity, adaptability, and generality of the proposed framework. \end{abstract}

\begin{IEEEkeywords}
CSI feedback, deep joint source-channel coding, autoencoder, deep learning.
\end{IEEEkeywords}

%
\IEEEpeerreviewmaketitle

\section{Introduction}
\label{Introduction}
\IEEEPARstart{P}{ursuing} higher data rates, which researchers in the communication community never stop, has promoted the evolution of the past five generations of mobile communication systems and will continuously give impetus to the development of the sixth generation (6G) communication system. Increasing the bandwidth is a feasible solution that has been adopted in the past five generations and can gain up to 1 GHz bandwidth. As the exploitation of the frequencies beyond mmWave in 6G \cite{saad2020a}, the available bandwidth further increases. However, the propagation losses become more severe at higher millimeter wave (mmWave) and terahertz (THz) frequencies \cite{wang2021a}, which will cause either more power consumption or a smaller coverage radius. 

Improving spectral efficiency (SE) is another effective method to achieve higher data rates. According to Shannon capacity theorem \cite{cover1999elements}, increasing the signal-to-noise ratio (SNR) can improve SE. Once the SNR stands at the high level, the power consumption for further improvement of SE is unaffordable, as the SNR appears inside of the logarithm of Shannon capacity and the derivative of Shannon capacity increases slower with the increase of the SNR. Multiple-input multiple-output (MIMO) identifies a way that improves the channel capacity at the outside of the logarithm by exploiting the spatial degrees of freedom. In addition, MIMO can potentially improve energy efficiency,  which naturally compensates for the severe power limitations of transmitting signals carried at higher frequencies.

As successors of MIMO, massive MIMO \cite{marzetta2016fundamentals} has become a critical technology for 5G, and ultra-massive MIMO is expected to be the key technology for 6G \cite{sarieddeen2019terahertz,yang20196g}. A typical massive MIMO scenario is that the base station (BS) with a large number of antennas can simultaneously serve multiple user equipments (UEs). To sufficiently exploit a large number of antennas, BS needs to have the knowledge of the instantaneous downlink channel state information (CSI). In time-division duplexing (TDD) mode, the BS estimates the uplink CSI through the pilot signals transmitted by the UE and then the downlink CSI can be inferred from the uplink CSI by using channel reciprocity. In frequency-division duplexing (FDD) mode, since the uplink and downlink work on different frequencies, channel reciprocity is no longer satisfied. A three-step interaction is applied to acquire the downlink CSI in FDD: (1) the BS first transmits pilot signals to the UE, then (2) the UE estimates the downlink CSI according to the pilot signals, and lastly (3) the UE feeds back the estimated downlink CSI to the BS. This CSI feedback mechanism inevitably occupies part of uplink resources and reduces the amount of remaining resources for data transmission. 

To reduce the feedback overhead, the codebook-based method which feeds back the codebook index instead of downlink CSI is  adopted in 5G (Type I and enhanced Type II) \cite{3gpp.38.214}. However, the increasing number of MIMO antennas from 5G to beyond 5G or 6G (e.g., from massive MIMO to ultra- massive MIMO) expands the codebook space and simultaneously aggravates the feedback overhead. It is key to compress CSI to an affordable quantity by exploiting the channel characteristics. The compressive sensing (CS) based methods transform the CSI to a sparse representation in some bases and feed back the sparse representation to reduce the feedback overhead \cite{daubechies2004iterative, metzler2016denoising, li2013efficient}. While the sparse assumption\textemdash the prerequisite of CS-based method\textemdash is not strictly satisfied in the practical system. Moreover, the iterative approaches in the existing CS-based reconstruction algorithms are usually time-consuming. 

\begin{figure}[!tb]
\centering
\subfigure[components in the SSCC transmitter]{
\label{sscc_encoder}
\includegraphics[width=0.95\columnwidth]{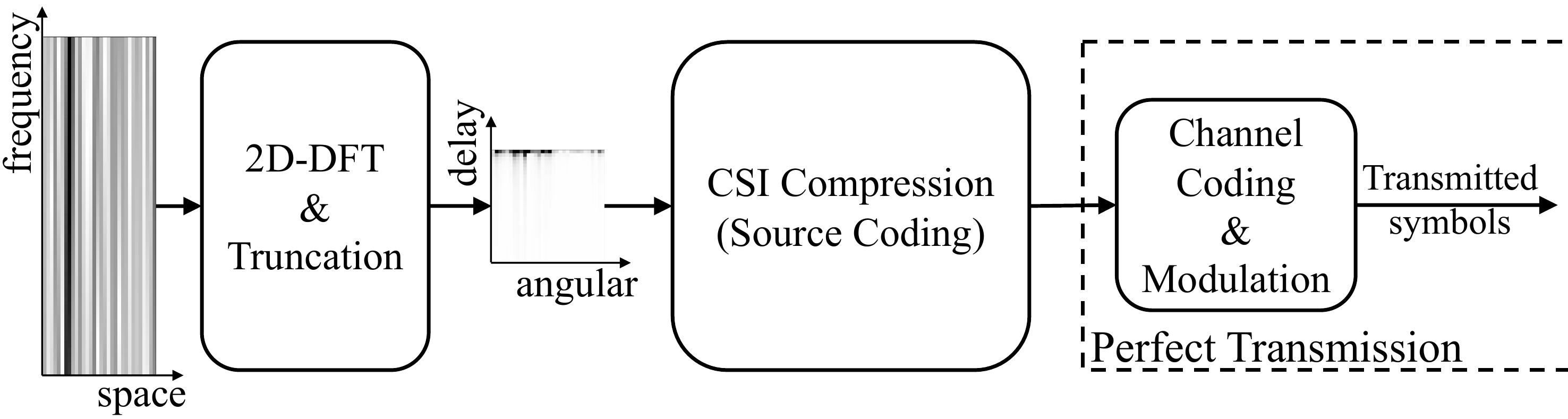}}
\subfigure[components in the DJSCC transmitter]{
\label{jscc_encoder}
\includegraphics[width=0.95\columnwidth]{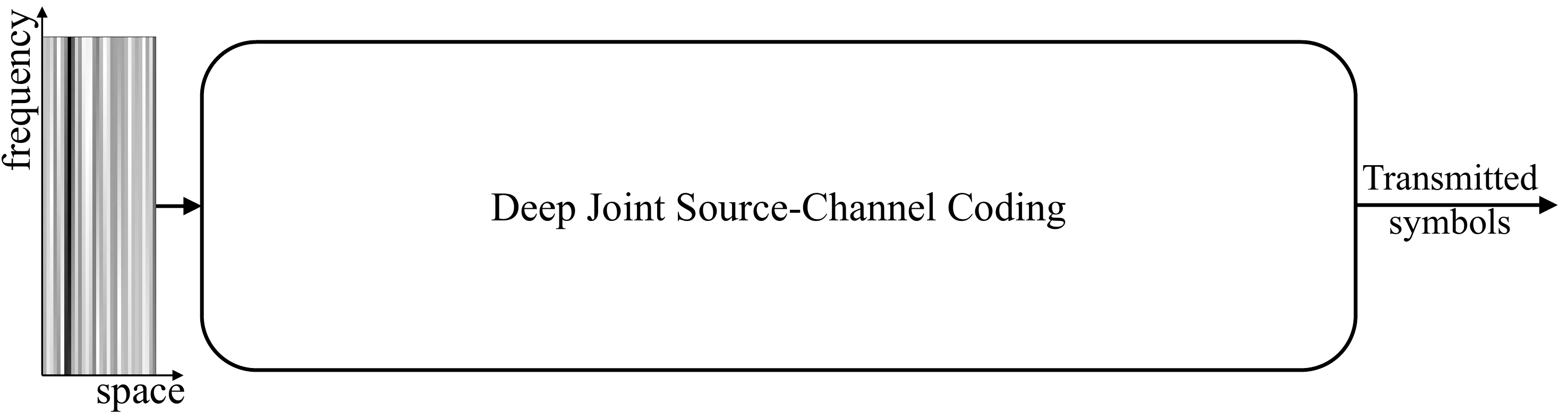}}
\caption{Overviews of the separate source-channel coding (SSCC) scheme and the deep joint source-channel coding (DJSCC) scheme. }\label{Fig:bw_valid}
\end{figure}

Till now, most of state-of-the-art methods in computer vision \cite{hassaballah2020deep}, natural language processing \cite{otter2020survey} and task-oriented communications \cite{shao2022learning} are based on deep learning (DL). These successes promote the exploitation of DL for the image compression task \cite{toderici2015variable,balle2016end,minnen2018joint,minnen2020channel}. According to the results revealed in \cite{toderici2015variable} and \cite{balle2016end}, the DL based image compression methods can achieve better performance than JPEG and JPEG2000, respectively. With the enhancements of hierarchical priors and latent residual prediction, the improved DL methods are even superior to Better Portable Graphics (BPG) in \cite{minnen2018joint} and \cite{minnen2020channel}. In fact, the CSI of massive MIMO can be regarded as a high-dimensional and low-rank image. The problem of the CSI feedback task can be seen as a problem of the image compression task. In this view, Wen et al. first proposed CsiNet, which uses the architecture of autoencoder in DL \cite{wen2018deep}. In that approach,, the CSI in the spatial-frequency domain is transformed to the angular-delay domain and continuously truncated to reduce CSI feedback overhead, which is shown as the 2D-DFT\&Truncation module in Fig. \ref{sscc_encoder}. The UE then compresses the truncated CSI in the angular-delay domain into codewords by the encoder and feeds back the codewords to the BS considering perfect transmission. The BS reconstructs the truncated CSI in the angular-delay domain from the codewords by the decoder. The proposed CsiNet outperforms the CS-based methods (e.g., LASSO \cite{daubechies2004iterative}, BM3D-AMP \cite{metzler2016denoising}, TVAL3 \cite{li2013efficient}) by a large margin. From then on, a series of research based on autoencoder architecture were proposed to explore the temporal correlation \cite{wang2018deep, lu2019mimo, li2020spatio}, improve the reconstruction accuracy \cite{guo2020convolutional, lu2020multi, hu2021mrfnet, chen2022high}, or balance the storage overhead, computational complexity and reconstruction accuracy \cite{cao2021lightweight, ji2021clnet, tang2021knowledge} in the truncated angular-delay domain. The aforementioned methods seem to have successfully solved the problem of the CSI feedback task. However, owing to the compressed CSI represented by 32-bit float values in the neural network, the number of feedback bits is 32 times the size of the output of the encoder, which is still a large amount for CSI feedback. To this end, novel quantization methods are proposed to further reduce CSI feedback overhead \cite{guo2020convolutional, lu2020bit, lu2021binary, mashhadi2021distributed}. 

Separation principles have guided classical communications system design in the past decades. Source coding, channel coding and modulation are designed separately. Most existing DL based methods for the CSI feedback task act as source coding, where the full CSI is the information to be compressed. The channel coding module and modulation module are assumed to be perfect, i.e., adaptable according to the feedback channel quality, and all feedback bits can be successfully transmitted. This scheme is regarded as the SSCC scheme shown in Fig. \ref{sscc_encoder}. However, there are several drawbacks. Firstly, this SSCC scheme has been demonstrated inferior to the joint source-channel coding (JSCC) scheme in the finite blocklength regime in theory \cite{kostina2013lossy}. Secondly, the SSCC scheme has ``cliff effect'' \cite{skoglund2006hybrid} in the real wireless scenario. That means the reconstruction quality of the CSI drops drastically, if the real feedback channel condition is worse than expected, and beyond the capability of the applied channel coding scheme. In this case, the recovered CSI at the BS is useless for the subsequent process. The JSCC scheme can provide a graceful performance degradation even the real channel condition becomes worse than the expected channel condition, which makes the recovered CSI still valuable for the subsequent process. Lastly, even though the hybrid automatic repeat request (HARQ) mechanism \cite{ahmed2021hybrid} can compensate for errors of channel decoding caused by channel condition mismatch, HARQ inevitably increases the additional feedback overhead and brings the latency problem for the CSI feedback task.

DL based JSCC (DJSCC) methods have shown reduced signal distortion in comparison to the SSCC methods with the use of the same amount of spectrum and energy resources \cite{farsad2018deep, bourtsoulatze2019deep, kurka2020deepjscc, kurka2021bandwidth, tung2021deepwive, xu2021wireless, xu2021deep}. For the text transmission task, the DJSCC method outperforms the traditional SSCC method in terms of word error rate under the erasure channel \cite{farsad2018deep}. For the image transmission task, the DJSCC methods proposed in \cite{bourtsoulatze2019deep} and \cite{kurka2020deepjscc} show better reconstruction performance than the JPEG/JPEG2000\&LDPC method and the BPG\&LDPC method, respectively. For the video transmission task, the proposed DJSCC method in \cite{tung2021deepwive} outperforms the H.264/H.265\&LDPC method in terms of the multi-scale structural similarity index measure (MS-SSIM). Inspired by resource assignment strategies in traditional JSCC \cite{sayood2000joint}, an attention based DJSCC method is proposed to deal with different SNR conditions with a single neural network, which balances the performance and the storage requirement in real wireless scenarios \cite{xu2021wireless}. Considering information privacy and confidentiality in using DJSCC, a novel source-channel coding method is proposed to protect the visual content of the plain image transmitted by the user \cite{xu2021deep}. The feasibility of combining DJSCC with orthogonal frequency division multiplexing (OFDM) system is demonstrated in \cite{yang2021deep}.


CSI feedback with noisy channels have been considered in \cite{guo2020convolutional, ye2020deep, mashhadi2020cnn}. The proposed method in  \cite{guo2020convolutional} is trained for a specific SNR and evaluated in different SNRs to show its robustness. The proposed method in \cite{ye2020deep} introduces a noise extraction unit for denoising the received codeword at the BS and a joint training strategy for improving the reconstruction performance. In \cite{mashhadi2020cnn}, a convolutional neural network (CNN)-based analog feedback method is proposed for the CSI in the spatial-frequency domain with a multi-carrier feedback channel. However, all the proposed methods in \cite{guo2020convolutional, ye2020deep, mashhadi2020cnn} need to train multiple models to combat with channel variations, which further aggravates the storage overhead both in the UE and the BS.

Different from the aforementioned methods for CSI feedback, in this paper, we design a general DJSCC framework for CSI feedback as shown in Fig. \ref{jscc_encoder}. The proposed framework focuses on efficiently executing the CSI feedback task over a noisy communication channel instead of considering accurately transmitting bits in traditional SSCC based communication systems. Compared with the SSCC scheme for CSI feedback, the proposed framework can improve the reconstruction performance of the CSI, overcome the problems of ``cliff effect'' and latency brought by the channel mismatch, and increase the uplink SE for CSI feedback. The major contributions are summarized as follows:

\begin{itemize}
\item[$\bullet$] We propose a unified DJSCC framework to deal with the CSI feedback task. Architectures of existing CSI compression networks can be exploited in the proposed framework. During the end-to-end training, the encoder network and the decoder network simultaneously learn from the CSI source and the wireless channel and, which leads to considerable performance gain.

\item[$\bullet$] We design two DL-based transform networks\textemdash one for converting the CSI from the spatial-frequency domain to a transform domain and the other for converting the CSI from a transform domain to the spatial-frequency domain\textemdash to reduce the distortion brought by dimensionality reduction in the existing CSI feedback methods.  

\item[$\bullet$] We introduce an SNR adaption strategy to match channel variations for DJSCC based CSI feedback. Instead of training multiple DJSCC networks with various SNRs, this strategy only needs to train a single DJSCC network with various SNRs. Moreover, it can save the storage space at the UE and the BS during network deployment in real wireless communication systems.

\end{itemize}

 The rest of this paper is organized as follows. Section II introduces the system model for the end-to-end CSI feedback. Section III overviews the separate source-channel coding framework based on the existing CSI methods and proposes a DJSCC based CSI feedback framework. In Section IV, the proposed method is evaluated to demonstrate its validity, adaptability, and generality. Finally, Section V concludes our work.
 
 \textbf{Notations:} Vectors and Matrices are denoted by boldface lower- and upper-case letters, respectively. Without special clarification, the vector is regarded as a column vector by default in this paper. The $i$-th element in the vector $\boldsymbol{\rm x}$ is represented by $x_i$. Similarly, $\boldsymbol{\rm x}_i$ (or $\boldsymbol{\rm x}_u^i$) denotes the $i$-th column vector of the matrix $\boldsymbol{\rm X}$ (or $\boldsymbol{\rm X}_u$). $\mathbb{R}$ and $\mathbb{C}$ denote the sets of real and complex numbers, respectively. $\|\cdot\|_2$ is the Euclidean norm. A circular symmetric Gaussian random vector $\boldsymbol{\rm x}$ with covariance matrix $\boldsymbol{\rm \Sigma}$ is denoted as  $\boldsymbol{\rm x} \sim\mathbb{CN}(0,\boldsymbol{\rm \Sigma})$. Finally, $(\cdot)^T$ and $(\cdot)^H$ denote the transpose operation and the conjugate transpose operation, respectively.

\section{System Model}

\label{system_model}
\begin{figure*}[!htb]
\centering
\includegraphics[width=2\columnwidth]{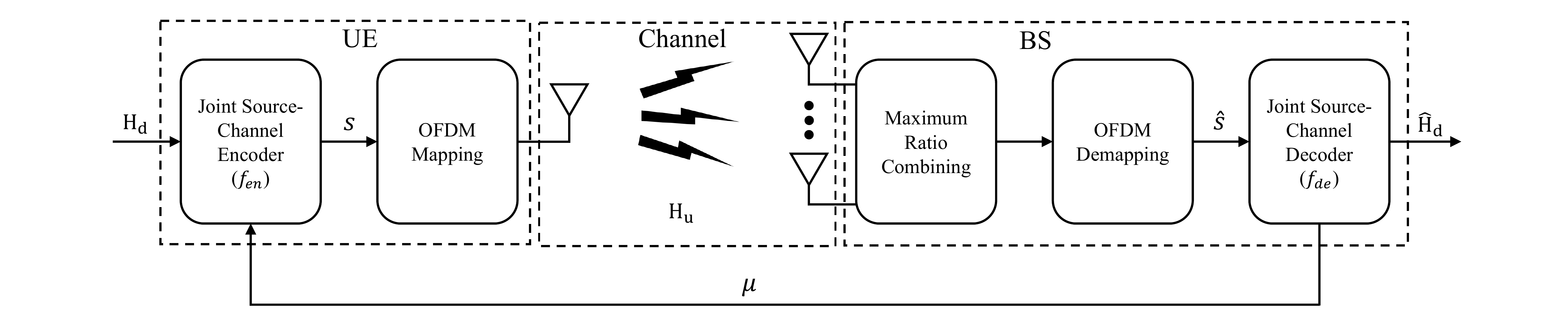}
\caption{The CSI feedback system with known SNR.}\label{Fig:system}
\end{figure*}

We consider a single cell FDD massive MIMO-OFDM system with $N_t\gg1$ antennas at a BS and a single antenna at a UE. Both the uplink and the downlink are over $N_c$ subcarriers. The downlink CSI and the uplink CSI in the spatial-frequency domain are represented by $\boldsymbol{\rm H}_d  \in \mathbb{C}^{N_c \times N_t}$ and $\boldsymbol{\rm H}_u  \in \mathbb{C}^{N_c \times N_t}$, respectively. 

The CSI feedback system with known SNR both in the UE and in the BS is shown in Fig.~\ref{Fig:system}. We assume that perfect uplink CSI and downlink CSI can be acquired after the channel estimation in the BS and in the UE, respectively. The joint source-channel encoder in the UE encodes the downlink CSI information $\boldsymbol{\rm H}_d$ and the known SNR $\mu \in \mathbb{R}$ into a vector $\boldsymbol{\rm s} \in \mathbb{C}^k$, which is expressed as: 
\begin{equation} 
\boldsymbol{\rm s}=f_{en}(\boldsymbol{\rm H}_d,\mu) \in \mathbb{C}^k,
\end{equation}
where $f_{en}: \mathbb{C}^{N_c \times N_t}\times \mathbb{R} \rightarrow \mathbb{C}^k$ represents the encoding function and $k \leq N_c\times N_t$ is the size of the encoded vector $\boldsymbol{\rm s}$. The uplink feedback channel SNR can be estimated at the BS and send to the UE in the downlink. The encoded vector $\boldsymbol{\rm s}=[s_1,s_2,\cdots, s_k]^T$ is mapped to $k$ subcarriers by OFDM mapping and then transmitted by the UE, where $s_i$ is the symbol carried on the $i$-th subcarrier. Assume the average power of a symbol over a subcarrier is $1$, then the encoded vector $\boldsymbol{\rm s}$ should be imposed a power normalization $ \frac{1}{k}\mathbb{E}(\boldsymbol{\rm ss^*})= 1$ to satisfy the power constraint at the UE.

The received $i$-th subcarrier of the feedback signal at the BS can be expressed as:
\begin{equation}
\boldsymbol{\rm y}_i= \boldsymbol{\rm h}_u^is_i+\boldsymbol{\rm z}_i,
\label{channel}
\end{equation}
where $\boldsymbol{\rm y}_i \in \mathbb{C}^{N_t}$ is the received vector containing the received symbols at the BS antennas over the $i$-th subcarrier, and $\boldsymbol{\rm z}_i \in \mathbb{C}^{N_t}$ consists of independent and identically distributed (i.i.d) samples with the distribution $\mathbb{CN}(0,\sigma^2\boldsymbol{\rm I})$. $\sigma^2$ represents the noise power.

After receiving multiple noisy replicas of $s_i$ $(i=1,2,\cdots,k)$ at the BS antennas, the recovered $\hat{s}_i$ over the $i$-th subcarrier is acquired by executing maximum ratio combining (MRC) at the BS, which is expressed as:
\begin{equation} 
\hat{s}_i= \boldsymbol{\rm w}^H_i\boldsymbol{\rm y}_i,
\label{mrc}
\end{equation}
For MRC, the combing vector $\boldsymbol{\rm w}_i$ is $\boldsymbol{\rm h}_u^i/\|\boldsymbol{\rm h}_u^i\|_2$. After the MRC process, the received symbols are converted to the vector $\boldsymbol{\rm \hat{s}}=[\hat{s}_1, \hat{s}_2,\cdots,\hat{s}_k]^T$, which represents the reconstruction of the vector $\boldsymbol{\rm s}$. The joint source-channel decoder at the BS uses a decoding function $f_{de}: \mathbb{C}^k \times \mathbb{R} \rightarrow \mathbb{R}^n$ to map $\boldsymbol{\rm \hat{s}}$ and the SNR $\mu$ to the recovered CSI $\boldsymbol{\rm \hat{H}}_d$, which is expressed as:

\begin{equation} 
\boldsymbol{\rm \hat{H}}_d=f_{de}(\boldsymbol{\hat{s}}, \mu) \in \mathbb{C}^{N_c\times N_t}.
\end{equation}
The feedback bandwidth is defined as $k$, which is the number of uplink subcarriers used for CSI feedback.

\begin{figure*}[!htb]
\centering
\includegraphics[width=2\columnwidth]{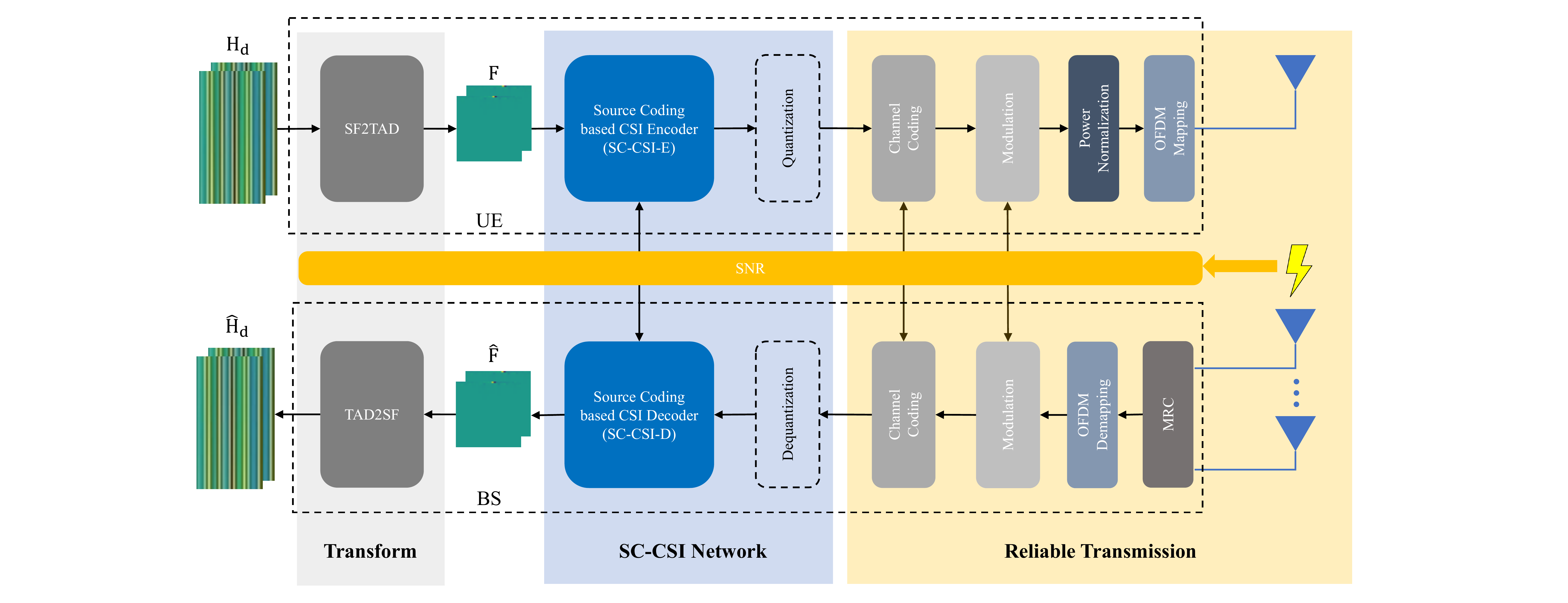}
\caption{The architecture of the SSCC based framework for the CSI feedback task. This architecture consists of three parts: (1) Transform, (2) SC-CSI network, and (3) Reliable transmission. The SF2TAD model converts the CSI from the spatial-frequency domain to the truncated angular-delay domain. Conversely, the TAD2SF model converts the CSI from truncated angular-delay domain to the spatial-frequency domain.
}\label{Fig:sscc}
\end{figure*}

\section{Deep Joint Source-Channel Coding based CSI Feedback}

In this section, we overview the SSCC based CSI feedback framework and propose a DJSCC based CSI feedback framework with SNR adaption. 
 
Based on the modular design principle (i.e., the SSCC scheme) in wireless communications, the existing DL based methods for the CSI feedback task\cite{wen2018deep, wang2018deep, lu2019mimo, li2020spatio, guo2020convolutional, lu2020multi, hu2021mrfnet, chen2022high, cao2021lightweight, ji2021clnet, tang2021knowledge, lu2020bit, lu2021binary} focus on CSI compression (i.e., source coding) with the promise of reliable transmission provided by appropriate channel coding and modulation as shown in Fig.~\ref{Fig:sscc}. Here, we call these existing CSI feedback methods source coding based CSI feedback (SC-CSI) methods. 

By exploiting the sparsity of the CSI information in the angular-delay domain, a 2D discrete Fourier transform (DFT) is executed at the UE to convert $\boldsymbol{\rm H}_d \in \mathbb{C}^{N_c\times N_t}$ from the spatial-frequency domain to the angular-delay domain, which is expressed as:
\begin{equation} 
\boldsymbol{\rm \widetilde{F}} = \boldsymbol{\rm I}_d\boldsymbol{\rm H}_d\boldsymbol{\rm D}_a,
\label{dft}
\end{equation}
where $\boldsymbol{\rm \widetilde{F}} \in \mathbb{C}^{N_c \times N_t}$ is the CSI information in the angular-delay domain, $\boldsymbol{\rm I}_d$ is an $N_c \times N_c$ inverse DFT (IDFT) matrix, and $\boldsymbol{\rm D}_a$ is an $N_t \times N_t$ DFT matrix. Due to the fact that most of the power among multiple paths  lies in a particularly limited period, the first $N_c^{\prime}$ rows which exhibit the distinct non-zero values are reserved as follows:
\begin{equation} 
\boldsymbol{\rm F} = Trun(\boldsymbol{\rm \widetilde{F}}) \in \mathbb{C}^{N_c^{\prime} \times N_t},
\label{trun}
\end{equation}
where $Trun(\cdot)$ represents a truncate operation in the delay dimension and $\boldsymbol{\rm F}$ is the truncated CSI in the angular-delay domain. These two stages are integrated as the spatial-frequency domain to truncated angular-delay domain (SF2TAD) module in Fig.~\ref{Fig:sscc}.

The SC-CSI encoder (SC-CSI-E) module at the UE encodes the truncated CSI $\boldsymbol{\rm F}$ to the vector $\boldsymbol{\rm c} \in \mathbb{R}^m$ by the encoding function $e_{\theta}: \mathbb{C}^{N_c^{\prime} \times N_t} \rightarrow \mathbb{R}^m$, which is expressed as:
\begin{equation} 
\boldsymbol{\rm c} = e_{\theta}(\boldsymbol{\rm F}) \in \mathbb{R}^{m},
\label{sc_encode}
\end{equation}
where $\theta$ is the parameter sets of the SC-CSI-E. In SC-CSI methods, a reliable transmission is assumed for transmitting the encoded vector $\boldsymbol{\rm c}$ from the UE to the BS without error. 
\begin{figure*}[!htb]
\centering
\includegraphics[width=2\columnwidth]{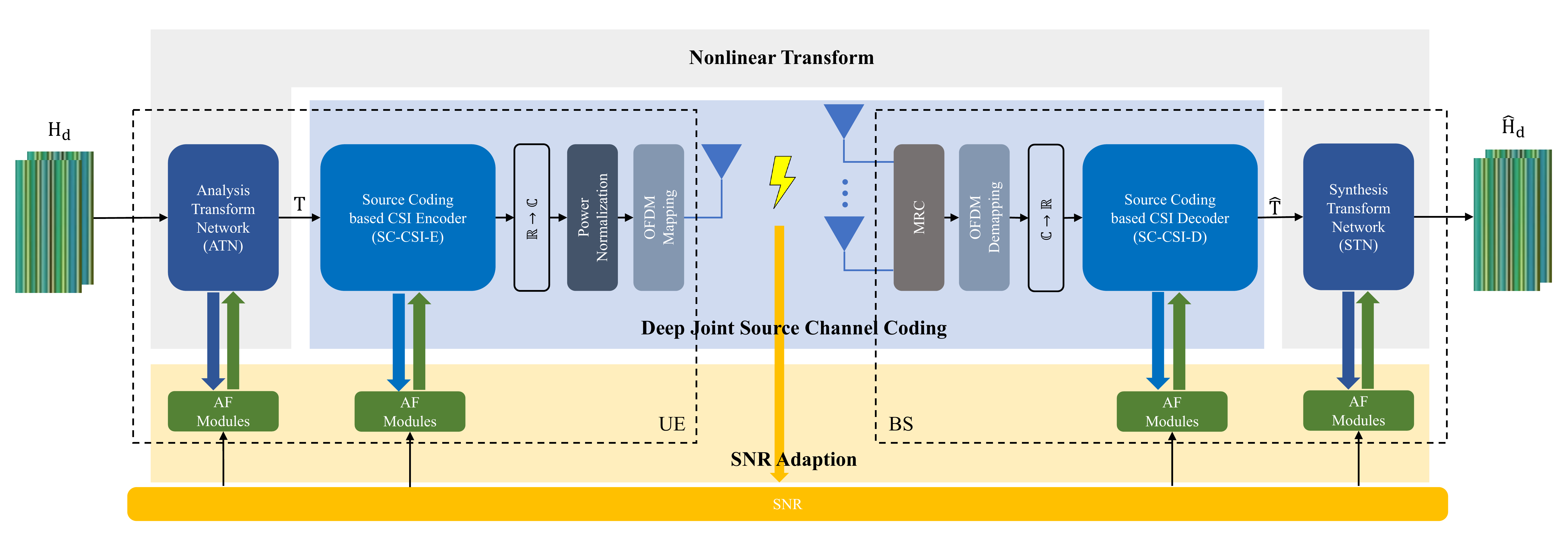}
\caption{The architecture of the DJSCC based framework for the CSI feedback task. This architecture consists of three parts: (1) Nonlinear transform, (2) Deep joint source channel coding, and (3) SNR adaption.}\label{Fig:adjscc}
\end{figure*}
The SC-CSI decoder (SC-CSI-D) module at the BS decodes the vector $\boldsymbol{\rm c}$ to an approximate reconstruction of the truncated CSI in the angular-delay domain $\hat{\boldsymbol{\rm F}} \in \mathbb{C}^{N_c^{\prime} \times N_t}$ by the function $d_{\phi}: \mathbb{R}^m \rightarrow \mathbb{C}^{N_c^{\prime} \times N_t}$ expressed as:
\begin{equation} 
\hat{\boldsymbol{\rm F}} = d_{\phi}(\boldsymbol{\rm c}) \in \mathbb{C}^{N_c^{\prime} \times N_t},
\label{sc_decode}
\end{equation}
where $\phi$ is the parameter set of the SC-CSI-D. The existing SC-CSI methods focus on the designs of the SC-CSI-E and SC-CSI-D networks, where the parameter sets $\{\theta,\phi\}$ are obtained by optimizing the following loss function $L_{sc}$:
\begin{equation} 
L_{sc}(\theta,\phi) =\frac{1}{T}\sum_{i=1}^T\|d_\phi(e_\theta(\boldsymbol{\rm F}^{(i)}))-\boldsymbol{\rm F}^{(i)}\|^2_2,
\label{loss_ad}
\end{equation}
where $T$ is the number of samples in the training dataset and $\boldsymbol{\rm F}^{(i)}$ represents the $i$-th sample in the training dataset in the truncated angular-delay domain.

After the decoding process at the BS, a zero matrix with the size $(N_c-N_c^{\prime})\times N_t$ is padded at the end of the recovered truncated CSI in the angular-delay domain $\hat{\boldsymbol{\rm F}}$ to acquire the recovered CSI in the angular-delay domain $\boldsymbol{\rm  \overline{F}}$, which is expressed as:
\begin{equation} 
\boldsymbol{\rm  \overline{F}} = Zeropadding(\boldsymbol{\rm \hat{F}}) \in \mathbb{C}^{N_c \times N_t}.
\end{equation}
The BS then converts the recovered CSI from the angular-delay domain to the spatial-frequency domain as follows:
\begin{equation} 
\boldsymbol{\rm \overline{H}}_d = \boldsymbol{\rm D}_f\boldsymbol{\rm \overline{F}}\boldsymbol{\rm I}_s,
\end{equation}
where $\boldsymbol{\rm \overline{H}}_d \in \mathbb{C}^{N_c \times N_t}$ is the recovered CSI in the spatial-frequency domain, $\boldsymbol{\rm D}_f$ is an $N_c \times N_c$ DFT matrix, and $\boldsymbol{\rm I}_s$ is a $N_t \times N_t$ IDFT matrix. These two stages are integrated as the truncated angular-delay domain to spatial-frequency domain (TAD2SF) module in Fig.~\ref{Fig:sscc}.

The assumption of the reliable transmission simplifies the design of the CSI feedback task based on the SSCC scheme. However, the performance of the SSCC scheme is inferior to that of the JSCC scheme in the finite blocklength regime in theory\cite{kostina2013lossy} and easily suffers from the ``cliff effect''. Once these SC-CSI methods are applied in these wireless communication systems, the adaptive modulation and coding (AMC) strategy should be employed to deal with the varying wireless channel and achieve a high throughput.

The DJSCC methods have been demonstrated to have better performance than SSCC methods for the text transmission task, the image transmission task, and the video transmission task\cite{farsad2018deep, bourtsoulatze2019deep, kurka2020deepjscc, kurka2021bandwidth, tung2021deepwive, xu2021wireless, xu2021deep}. In this section, we propose a DJSCC based framework with SNR adaption for the CSI feedback task (ADJSCC-CSI) as shown in Fig.~\ref{Fig:adjscc}, which can exploit the experiences acquired in the existing SC-CSI methods, improve the CSI reconstruction performance, overcome the ``cliff effect'', and increase the SE for CSI feedback. There are three parts in the ADJSCC-CSI framework as shown in Fig.~\ref{Fig:adjscc}, including nonlinear transform, deep joint source-channel coding and SNR adaption.
	
\subsection{Nonlinear Transform} 
\begin{figure}[!tb]
\centering
\includegraphics[width=1\columnwidth]{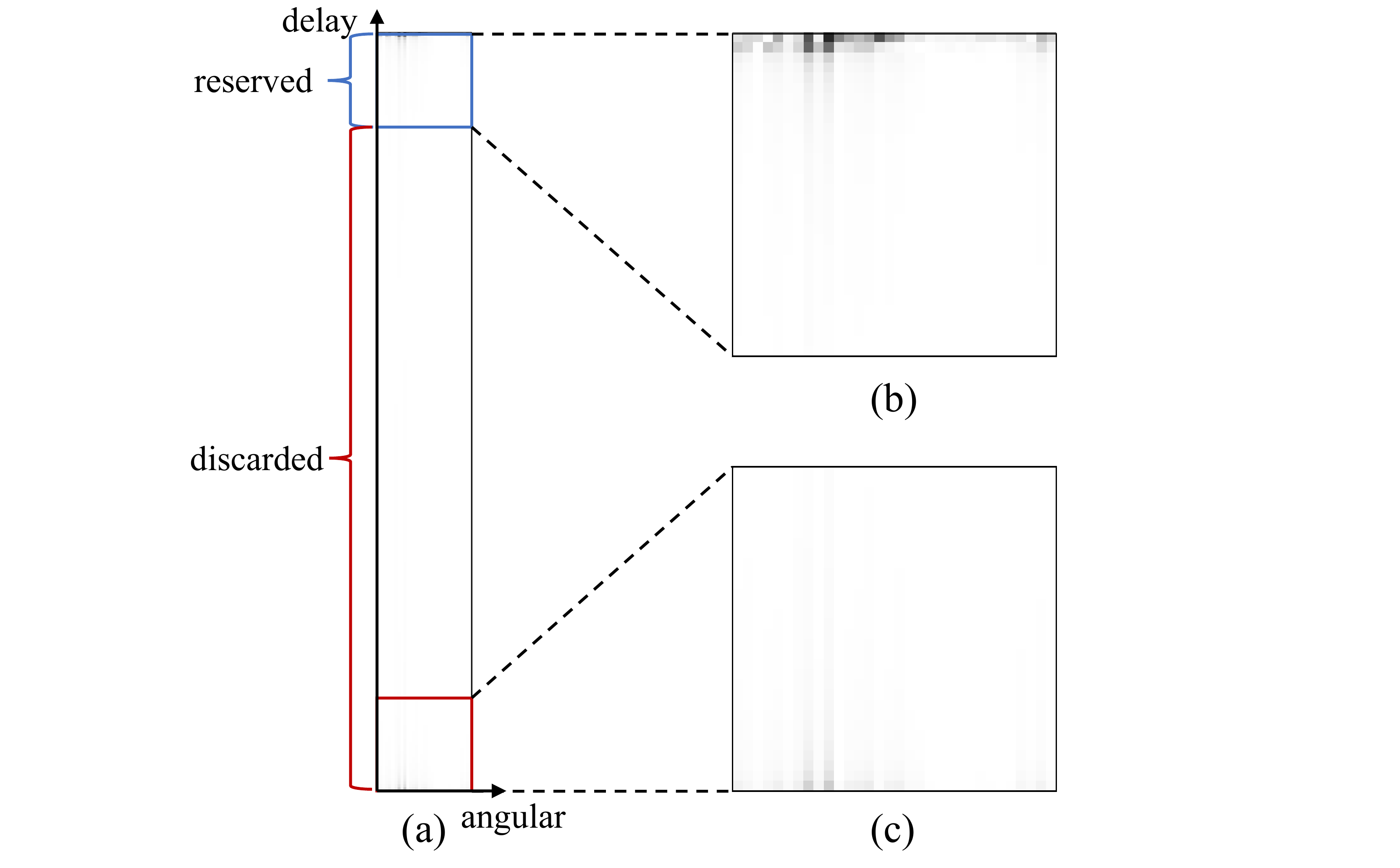}
\caption{Visualized images of the CSI sample in the angular-delay domain. The CSI sample is from the dataset generated by QuaDRiGa in Section \ref{section_4}. (a) The whole part of the CSI in the angular-delay domain, (b) The reserved part of the CSI in the angular-delay domain for SC-CSI methods, (c) The discarded part of the CSI in the angular-delay domain for SC-CSI methods.}\label{Fig:csi}
\end{figure}

According to the sparsity assumption of CSI in the angular-delay domain, the CSI information in the spatial-frequency domain $\boldsymbol{\rm H}_d$ is first converted to the angular-delay domain $\boldsymbol{\rm F}$ by Eq. \eqref{dft} and then truncated by Eq. \eqref{trun} for dimensionality reduction in SC-CSI methods. Fig.~\ref{Fig:csi}(a) shows the CSI visualization in the angular domain. Fig.~\ref{Fig:csi}(b) shows the reserved part of the CSI for SC-CSI methods. The element with light color means the value of the element is close to zero, whereas the element with dark color means the value of the element is large. 
Indeed, Fig.~\ref{Fig:csi}(b) contains most of non-zero elements, which is then compressed in SC-CSI methods. However, some useful information about the CSI shown in Fig.~\ref{Fig:csi}(c) is discarded and cannot be compensated in the subsequent process, which will lead to some distortion of the recovered CSI in SC-CSI methods. Therefore, it is desired to modify the imperfect truncation adopted in SC-CSI methods.

In this paper, we exploit nonlinear transforms for CSI source. In image compression domain, the DL based non-linear transform \cite{balle2016end} has been demonstrated superior to the the discrete cosine transform in JPEG \cite{wallace1992jpeg} and the wavelet transform in JPEG2000 \cite{rabbani2002jpeg2000}.  The nonlinear transforms are implied by using the analysis transform network (ATN) at the UE:
\begin{equation} 
\boldsymbol{\rm  T} = A_{\alpha}(\boldsymbol{\rm H}_d) \in \mathbb{C}^{N_c^{\prime} \times N_t},
\end{equation}
 and the synthesis transform network (STN) at the BS:
\begin{equation} 
\boldsymbol{\rm  \hat{H}}_d = S_{\beta}(\boldsymbol{\rm \hat{T}}) \in \mathbb{C}^{N_c \times N_t},
\end{equation}
where $\boldsymbol{\rm T} \in \mathbb{C}^{N_c^{\prime} \times N_t}$ is the transformed CSI at the UE,  $\boldsymbol{\rm \hat{T}} \in \mathbb{C}^{N_c^{\prime} \times N_t}$ and $\boldsymbol{\rm \hat{H}}_d \in \mathbb{C}^{N_c \times N_t}$ are the recovered CSI in the transform domain and the recovered CSI in the spatial-frequency domain at the BS, respectively. To simplify the notation and enable fair comparison with SC-CSI methods, the CSI in transform domain $\boldsymbol{\rm T} $ and the recovered CSI in the transform domain $\boldsymbol{\rm \hat{T}}$ have the same size with the truncated CSI in the angular-delay domain $\boldsymbol{\rm F}$. With an end-to-end training strategy, the ATN and the STN can learn more effective transforms than 2D DFT used in SC-CSI methods.

\subsection{Deep Joint Source-Channel Coding} 

The power of DJSCC comes from both the design of the neural network and the principle of JSCC. A typical DJSCC structure consisting of the joint source-channel encoder, noisy channel, and the joint source-channel decoder is introduced in \cite{bourtsoulatze2019deep}. The DJSCC encoder maps the input to a low dimensional vector with complex values at first. The real part and the imaginary part of the low dimensional vector is regarded as the I and Q components of the modulated symbols and directly transmitted through the noisy channel. Then the DJSCC decoder receives the noisy symbols and maps the corrupted symbols to an estimation of the original input. A differentiable channel transfer function is necessary for supporting back propagation in the training stage. During the end-to-end training stage, the DJSCC learns effective parameters for the encoder and the decoder by simultaneously perceiving the input source and the noisy channel.

As complex encoded symbols are required in the DJSCC encoder, the output of the SC-CSI-E is first converted to the complex values as follows:
\begin{equation} 
\boldsymbol{\rm  s} = R2C(\boldsymbol{\rm  c}) \in \mathbb{C}^k,
\label{r2c}
\end{equation}
where $\boldsymbol{\rm  c} \in \mathbb{R}^{2k}$ is the output of the SC-CSI-E,  $R2C(\cdot)$ represents the function converting the real vector $\boldsymbol{\rm  c}$ to the $k$ dimensional vector $\boldsymbol{\rm s}$. According to the power constraint, a power normalization function $PowerNorm(\cdot)$ is applied at the UE:
 \begin{equation} 
\boldsymbol{\rm  s} = PowerNorm(\boldsymbol{\rm  s}) \in \mathbb{C}^k.
\label{pn}
\end{equation}
In this paper, we adopt average power constraint. Then the symbols are mapped to subcarriers and transmitted by the UE. At the BS side, after the MRC and OFDM demapping, the received complex symbols $\boldsymbol{\rm \hat{s}} \in \mathbb{C}^k$ is converted  to real values as follows:
\begin{equation} 
\boldsymbol{\rm  \hat{c}} = C2R(\boldsymbol{\rm  \hat{s}}) \in \mathbb{R}^{2k},
\label{c2r}
\end{equation}
 where $C2R(\cdot)$ represents the function that converts the vector $\boldsymbol{\rm \hat{s}}$ to the $2k$ dimensional vector $\boldsymbol{\rm  \hat{c}}$. Then the SC-CSI-D decodes the vector $\boldsymbol{\rm \hat{c}}$. Based on these modifications, the architectures in existing SC-CSI method can be can be exploited in the proposed DJSCC framework. These conversions endow the encoding network and the decoding network ability to explore the knowledge from both the source and channel by the end-to-end training, which improves the reconstruction performance of the CSI and overcomes the problem of ``cliff effect'' in the SSCC scheme.

\subsection{SNR Adaption} 
\label{subsection_3C}
In the training stage, the DJSCC based methods are trained under the specific channel condition to effectively explore the channel characteristics \cite{bourtsoulatze2019deep, kurka2020deepjscc, kurka2021bandwidth}. Multiple networks trained under different channel conditions are then deployed to match the real channel conditions. The large storage overhead limits its applicability in resource-constrained devices. Inspired by resource assignment strategy in traditional JSCC, our previous work \cite{xu2021wireless} successfully designs a single network to cover different channel conditions by introducing SNR feedback from the decoder to the encoder. Here, we extend our previous work to deal with the CSI feedback task under fading channels. The core module in \cite{xu2021wireless} is Attention Feature (AF) Module, which is a plug and play module. The AF Module can be employed for main layers (e.g., the convolution layer and the transposed convolution layer) of the neural network. Specifically, The inputs of the AF module are the SNR and the output of the main layer, and the output of the AF module is the scaled parameters multiply the output of the main layer. The function of the AF Module is to produce a sequence of scaling parameters to reassign the importance of the features in this layer. 

Combining with the AF Modules, the ATN and the SC-CSI-E at the UE are improved as follows:
\begin{equation} 
\boldsymbol{\rm T} = AF_\gamma (A_{\alpha}(\boldsymbol{\rm H}_d), \mu) \in \mathbb{C}^{N_c^{\prime} \times N_t},
\end{equation}
\begin{equation} 
\boldsymbol{\rm c} = AF_\psi(e_{\theta}(\boldsymbol{\rm T}), \mu) \in \mathbb{R}^{m},
\end{equation}
where $AF_\gamma(\cdot)$ and $AF_\psi(\cdot)$ denote that the AF Modules with parameter sets $\gamma$ and $\psi$ are placed after each main layer in the ATN and the SC-CSI-E as reference \cite{Zhang_2018_ECCV}, respectively. The SNR $\mu \in \mathbb{R}$ is the feedback SNR from the BS to the UE. The output vector $\boldsymbol{\rm c}$ with real values is used to generate the normalized symbols with complex values $\boldsymbol{\rm s}$ according to Eq. \eqref{r2c} and Eq. \eqref{pn}. The complex vector $\boldsymbol{\rm s}$ is mapped to subcarriers and transmitted through the fading channel, which is modeled in Eq. \eqref{channel}. The received symbols at the BS antennas are combined by MRC as in Eq. \eqref{mrc} and then demapped to the corrupted complex symbols $\boldsymbol{\rm \hat{s}}$. The recovered vector with real values $\boldsymbol{\rm \hat{c}}$ is converted according to Eq. \eqref{c2r}.

At the BS, the SC-CSI-D and the STN are combined with AF Modules expressed as:
\begin{equation} 
\hat{\boldsymbol{\rm T}} = AF_\rho(d_{\phi}(\boldsymbol{\rm c}), \mu) \in \mathbb{C}^{N_c^{\prime} \times N_t},
\end{equation}
\begin{equation} 
\boldsymbol{\rm  \hat{H}}_d = AF_\tau(S_{\beta}(\boldsymbol{\rm \hat{T}}), \mu) \in \mathbb{C}^{N_c \times N_t},
\end{equation}
where $AF_\rho(\cdot)$ and $AF_\tau(\cdot)$ denote that the AF Modules with parameter sets $\rho$ and $\tau$ are placed after each main layer in the SC-CSI-D and the STN, respectively.

Different from existing SC-CSI methods considering the loss function Eq. \eqref{loss_ad} in the angular-delay domain, we optimize the parameter sets $\Theta=\{\alpha, \theta, \phi, \beta, \gamma, \psi, \rho, \tau\}$ by minimizing the distortions under a certain bandwidth $k$ as follows:
\begin{equation} 
\Theta^* = \mathop{\arg\min}\limits_{\Theta} \mathbb{E}_{p(\mu)}\frac{1}{T}\sum_{i=1}^T\|\boldsymbol{\rm H}_d^{(i)}-\boldsymbol{\rm \hat{H}}_d^{(i)}\|^2_2,
\end{equation}
where $\Theta^*$ is the optimal parameter set, $p(\mu)$ represents the probability distribution of the SNR, $\boldsymbol{\rm H}_d^{(i)}$ represents the $i$-th sample of the training dataset in the spatial-frequency domain, and $\boldsymbol{\rm \hat{H}}_d^{(i)}$ represents the reconstruction of $\boldsymbol{\rm H}_d^{(i)}$ at the BS. During the training stage, the proposed ADJSCC-CSI for the CSI feedback task can learn: (1) a pair of transformations separately from the spatial-frequency domain to the transformed domain at the UE and from the transformed domain to the spatial-frequency domain at the BS, (2) a pair of encoder and decoder built upon the SC-CSI methods which learn from both the source and the channel, and (3) an end-to-end network which can adapt against the channel variation.

\section{Experimental Results}
\label{section_4}
The proposed ADJSCC-CSI is a general framework that can be employed to upgrade many SC-CSI methods to DJSCC based methods. We first demonstrate its effectiveness by using a concrete scheme in this section. Next, we evaluate the ability of SNR adaption under this concrete scheme. Finally, we illustrate the generality of ADJSCC-CSI by exploiting different architectures in other SC-CSI methods.

The uplink CSI and downlink CSI are generated by QuaDRiGa \cite{quadriga} according to the 3rd Generation Partnership Project (3GPP) TR 38.901 \cite{TR38.901}. We create an open indoor scenario in which the downlink center frequency is 5.2 GHz and the uplink center frequency is 5.4 GHz. The BS is positioned at the center of a square area with the size of $20m \times 20m$. A uniform linear array (ULA) with half-wavelength antenna space is deployed at the BS. The number of the antennas at the BS is $N_t=32$ and the number of the antennas at the UE is $N_r=1$. Both the antennas in the BS and in the UE are omnidirectional. The heights of the BS and the UE are $3m$ and $1.5m$, respectively. The uplink and the downlink both have $N_c=256$ subcarriers. The training, validation, and testing sets contain 100,000, 30,000, and 20,000 sample pairs, respectively. A sample pair contains a downlink CSI sample and an uplink CSI sample. Following the setting in the existing SC-CSI methods \cite{wen2018deep, wang2018deep, lu2019mimo, li2020spatio, guo2020convolutional, lu2020multi, hu2021mrfnet, chen2022high, cao2021lightweight, ji2021clnet, tang2021knowledge, lu2020bit, lu2021binary}, $N_c^{\prime}$ is set to 32.

\begin{figure}[!tb]
\centering
\subfigure[ATN combined with AF Modules]{
\label{af_atn}
\includegraphics[width=0.9\columnwidth]{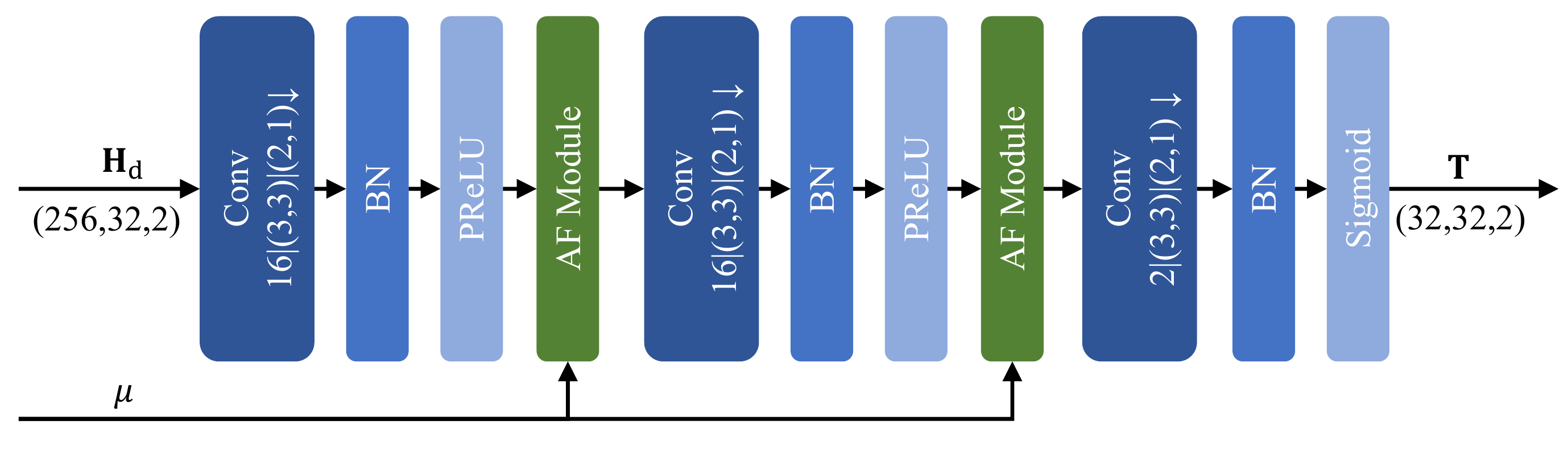}}
\subfigure[STN combined with AF Modules]{
\label{af_stn}
\includegraphics[width=0.9\columnwidth]{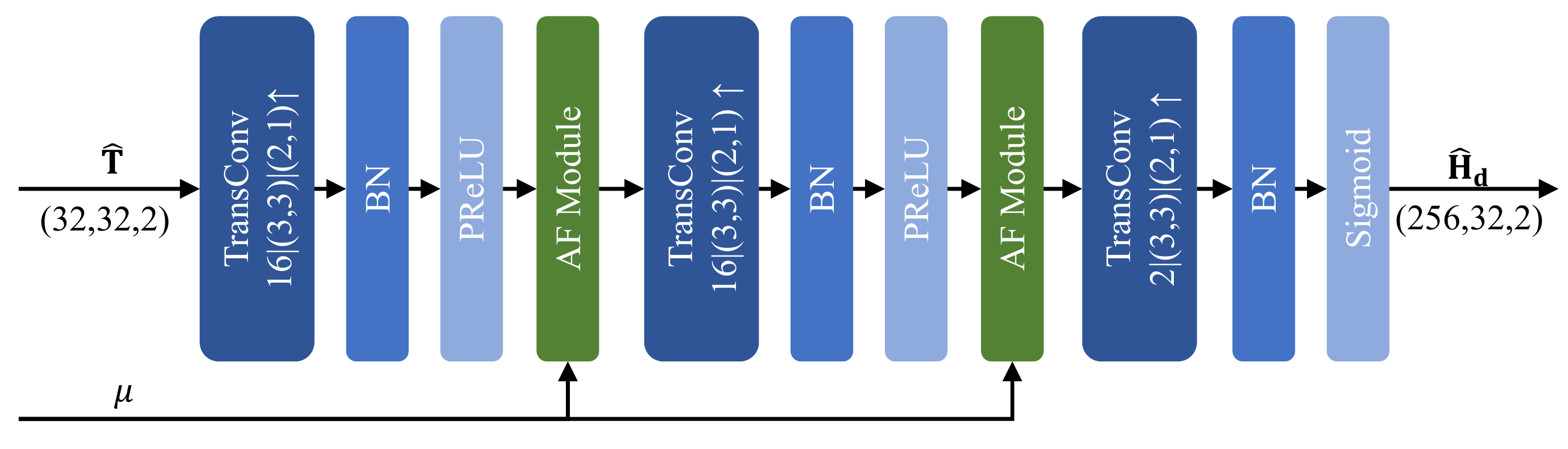}}
\caption{The architectures of the transform network combined with AF Modules. The notation $C|(F_1,F_2)|(S_1,S_2)\downarrow$ in the convolution layer denotes that it has C filters with size $F_1 \times F_2$. $S_1$ is the vertical steps for stride operation and $S_2$ is the horizontal steps for stride operation. $\uparrow$ and $\downarrow$ represents the stride up and down operation, respectively. 
}\label{Fig:af_transform}
\end{figure}

Fig.~\ref{Fig:af_transform} illustrates the networks of the ATN and the STN combined with AF Modules. The ATN combined with AF Modules shown in Fig.~\ref{af_atn} consists of three convolution layers, each of which is followed by a batch normalization layer and a PReLU \cite{he2015delving}  activation expected the last one. The sigmoid activation follows the last convolution layer. The AF Modules are inserted after the first two activation functions. The notation $C|(F_1,F_2)|(S_1,S_2)\downarrow$ in the convolution layer denotes that it has $C$ filters with size $F_1 \times F_2$. $\downarrow$ represents the stride down operation. $S_1$ and $S_2$ are the vertical steps and the horizontal steps, respectively. Due to the unsymmetrical steps the stride down operation adopts, the CSI information $\boldsymbol{\rm H}_d$ of a rectangle shape can be successfully converted to the transformed CSI $\boldsymbol{\rm T}$ of a square shape, which matches the truncated CSI size in the existing SC-CSI methods. The STN combined with AF Modules is shown in Fig.~\ref{af_stn}. It consists of three transposed convolution layers. Each of the first two transposed convolution layers is followed by a batch normal layer, a PReLU activation and an AF Module. The activation of the last transposed layer is also sigmoid. A more sophisticated network structure than the proposed one for the ATN and the STN may improve the CSI reconstruction performance. However, in this paper, we focus on the mechanism design rather than the network design.

Besides a new network architecture CsiNet+ for CSI feedback, \cite{guo2020convolutional} proposes a concrete quantization method for the real values of the CsiNet+ encoder. The proposed quantized method provides a more effective representation for the encoded output than its unquantized version in the view of bit compression. We first adopt the CsiNet+ encoder and the CsiNet+ decoder as the SC-CSI-E and the SC-CSI-D in Fig.~\ref{Fig:adjscc}, respectively. Consistently with the work \cite{guo2020convolutional}, the batch size is set to 200. Adam optimizer is first initialized with a learning rate of $10^{-3}$. When the loss does not decrease in 20 epochs, the learning rate will be decayed by half. The lower bound of the learning rate is set to $10^{-4}$. The training epochs are set to 500 for the network convergence. Tensorflow \cite{abadi2016tensorflow} and its high-level Keras are used to implement the following experiments.
All of our experiments are performed on a Linux server with four Hygon 7151 16-core CPUs and sixteen NVIDIA RTX A3000 GPU. Each experiment runs on four CPU cores and a GPU. Consistently with existing SC-CSI methods for the CSI feedback task, normalized mean square error (NMSE) is utilized to evaluate CSI reconstruction performance, which is expressed as: 
\begin{equation} 
\rm  NMSE = \mathbb{E}{\|\boldsymbol{\rm H}_d- \boldsymbol{\rm \hat{H}}_d\|_2^2/\|\boldsymbol{\rm H}_d\|_2^2}.
\end{equation}
The lower NMSE reveals the better performance of CSI feedback. 

\subsection{ADJSCC-CSI Validity Experiments}
\label{subsection_4A}
\begin{table}[!tb] 
\renewcommand{\arraystretch}{1.3} 
\caption{UCI Encoding Scheme} 
\label{Table:UCI}
\centering
\begin{tabular}{c|c|c|c}
\hline
\bfseries Length (L) & \bfseries Segmentation & \bfseries CRC Bits & \bfseries Encoding scheme\\ \hline
1 & N/A & N/A & Repetition\\ \hline
2 & N/A & N/A & Simplex\\ \hline
3-11 & N/A & N/A & Reed-Muller\\ \hline
12-19 & N/A & 6 & Parity-check Polar\\ \hline
20-1706 & L $\geq$ 1013 & 11 & Polar \\ \hline
\end{tabular}
\end{table}

The proposed framework built upon CsiNet+ for the CSI feedback task is named as ADJSCC-CsiNet+. The ADJSCC-CsiNet+ is trained under the uniform distribution of $\rm SNR_{train}$ from $-10$ dB to $10$ dB. To demonstrate the validity of the ADJSCC-CsiNet+, the SSCC scheme as shown in Fig.~\ref{Fig:sscc}, which is built upon CsiNet+ and 5G NR uplink control Information (UCI) transportation \cite{3gpp.38.212}, is used as the compared method. UCI encoding and decoding are available for channel coding, e.g., Polar coding, of small block lengths. The concrete UCI encoding scheme depends on the length of input bits shown in Table \ref{Table:UCI}. The modulation scheme of the UCI contains BPSK, QPSK, 16QAM, 64QAM, and 256QAM. To fairly compare with the ADJSCC-CsiNet+, source coding rate, channel coding rate, and modulation scheme should be adjusted to match the feedback bandwidth.

\begin{figure}[!tb]
\centering
\includegraphics[width=1\columnwidth]{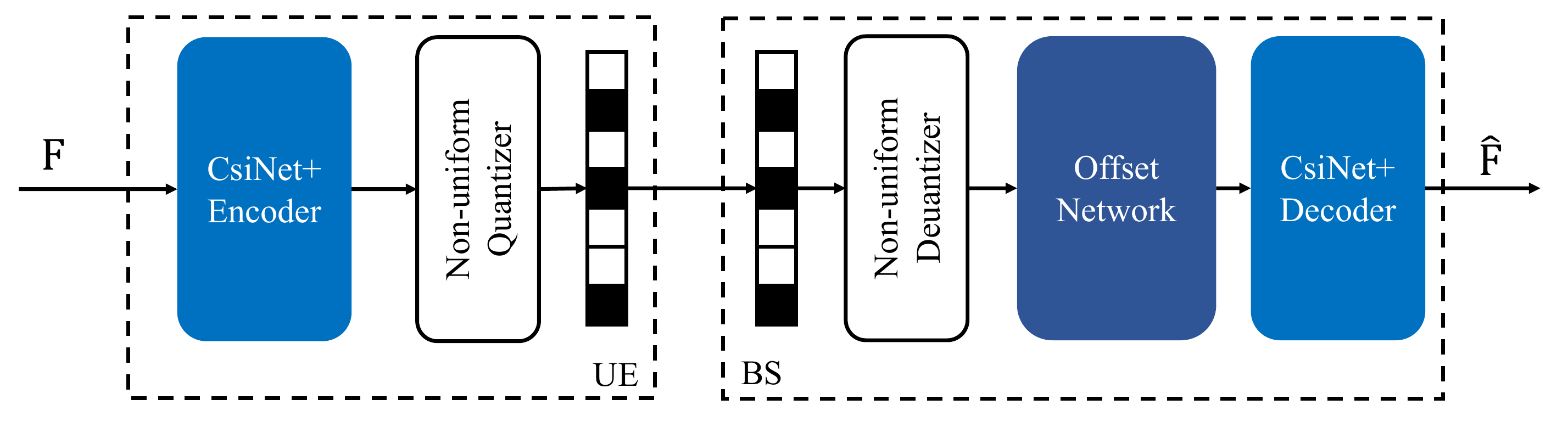}
\caption{Proposed bit-level CsiNet+ framework proposed by \cite{guo2020convolutional}. The original CSI in the angular-delay domain is first compressed at the CsiNet+ encoder and then quantized by the non-uniform quantizer to generate a bitstream in the UE. In the BS, the received bitstream are first dequantized and then fed into offset network for quantization distortion compensation. Finally, the CSI in the angular-delay domain is recovered by the CsiNet+ decoder.
}\label{Fig:CsiNet_plus_q}
\end{figure}

The framework of bit-level CsiNet+ network is shown in Fig.~\ref{Fig:CsiNet_plus_q}. The source coding rate varies for different quantized levels of the CsiNet+. The training strategy of the quantized CsiNet+ contains the following three steps: (1) The CsiNet+ without quantization is trained by an end-to-end approach in the truncated angular-delay domain; (2) The CsiNet+ encoder, non-uniform quantizer, and the non-uniform dequantizer are combined to generate the training dataset, which is then used to train the offset network for quantization compensation by minimizing the mean square error (MSE) between the training dataset and the output of the offset network; (3) The parameters of the CsiNet+ encoder are fixed and the parameters of the offset network and the CsiNet+ decoder are fine tuned to further improve the reconstruction performance. More training Details about bit-level CsiNet+ can be found in \cite{guo2020convolutional}.

\begin{figure}[!tb]
\centering
\subfigure[]{
\label{bw16_valid}
\includegraphics[width=0.9\columnwidth]{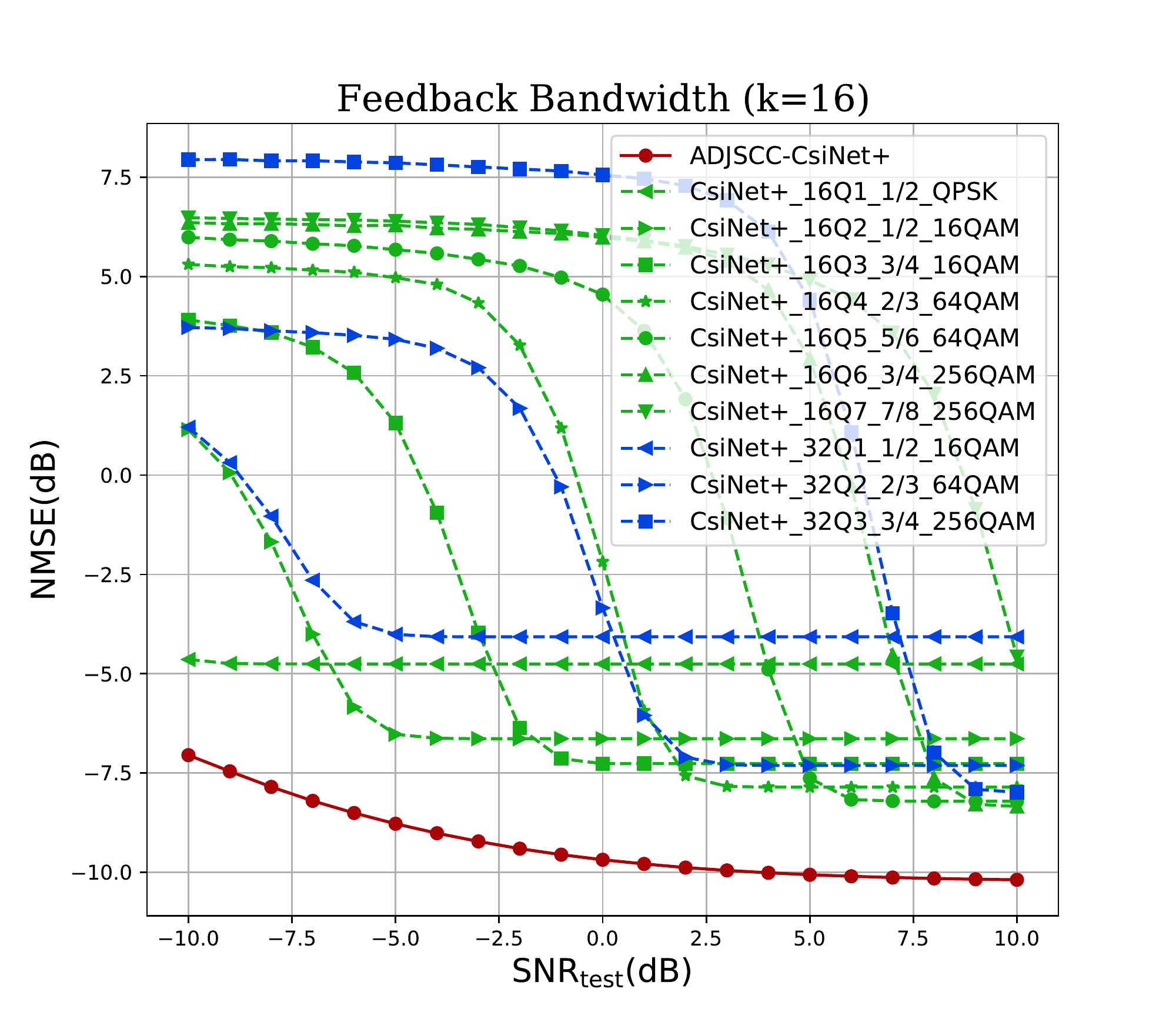}}
\subfigure[]{
\label{bw32_valid}
\includegraphics[width=0.9\columnwidth]{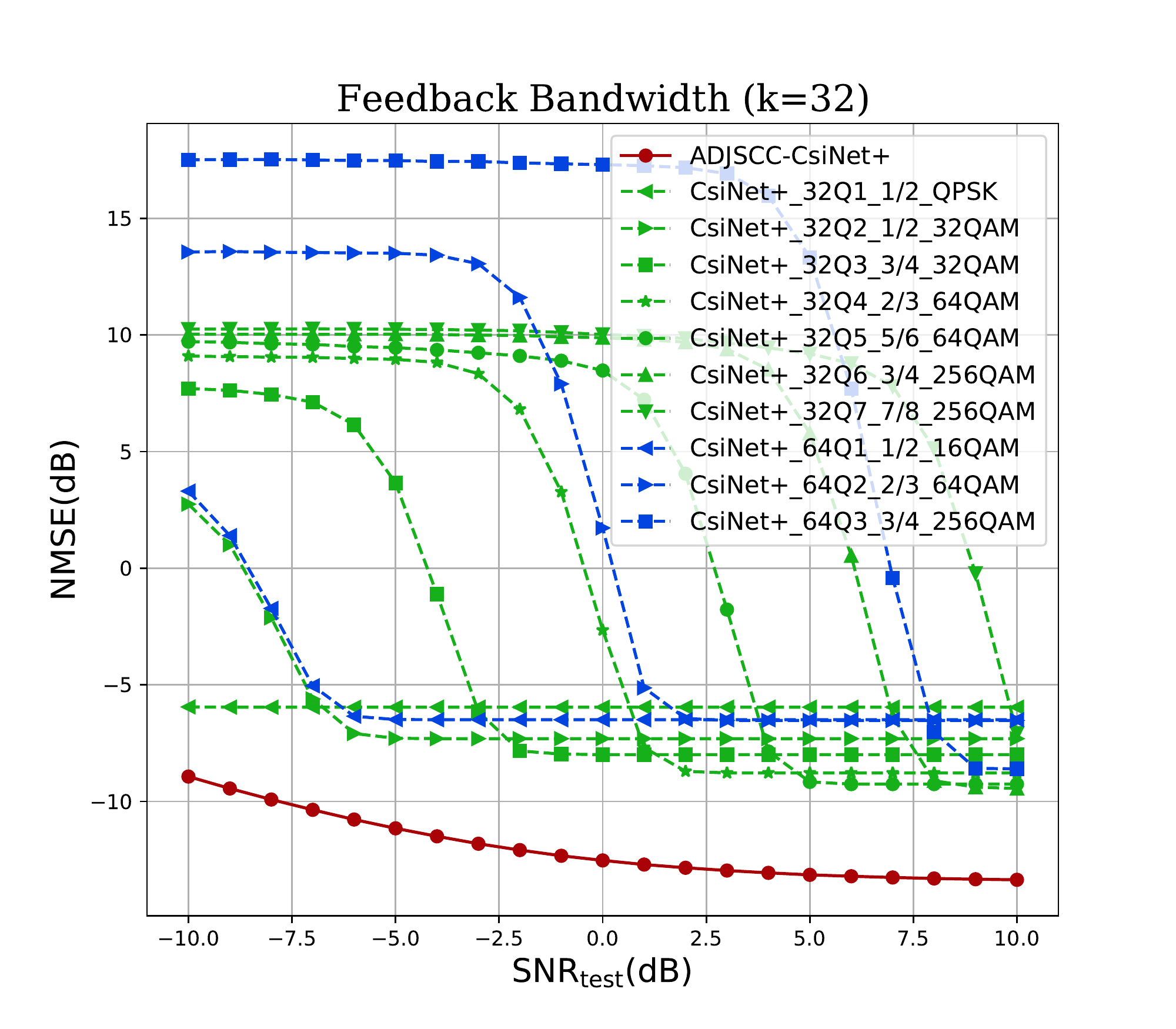}}
\caption{Performance comparison of ADJSCC-CsiNet+ with SSCC based CsiNet+. (a) k = 16 and (b) k=32. The SSCC based CsiNet+ consists of the CsiNet+ as the source coding scheme and 5GNR uplink control Information (UCI) scheme the as channel coding and modulation scheme. The label ``CsiNet+\_16Q2\_ 1/2\_16QAM'' means the CsiNet+ compresses the truncated CSI to a 16-dimensional vector, each of which is quantized to 2 bits and the UCI adopts 1/2 channel coding rate and 16QAM modulation scheme.}\label{Fig:bw_valid}
\end{figure}

Fig.~\ref{Fig:bw_valid} compares the ADJSCC-CsiNet+ method with the SSCC methods (i.e., CsiNet+ and UCI transmission) at the feedback bandwidth $k=16, 32$, respectively. The evaluation of CSI reconstruction performance is in the spatial-frequency domain. The label ``CsiNet+\_16Q2\_1/2\_16QAM'' represents that the SSCC method adopts the CsiNet+ as the source coding scheme. The CsiNet+ encoder compresses the truncated CSI to a 16-dimensional vector, each of which is quantized to 2 bits. Due to the length of the bitstream being larger than 20, the UCI then uses Polar coding as the channel coding scheme with channel coding rate 1/2 and the 16QAM modulation scheme to map the bitstream to the transmitted symbols. In Fig.~\ref{bw16_valid} with $k=16$, the performance of the ADJSCC-CsiNet+ is better than the performance of any SSCC method. In the SSCC method, the fixed source coding rate, channel coding rate, and modulation scheme are designed for the specific SNR. If the channel condition is worse than the target channel condition, the outbreak of channel decoding errors causes the reconstruction quality to drop drastically; If the channel condition is better than the target channel condition, the reconstruction quality remains the same. This is the ``cliff effect'' in SSCC methods. For instance, the SSCC method ``CsiNet+\_16Q4\_2/3\_64QAM'' suffers from drastic degradation when the channel SNR decreases from $2$ dB to $-10$ dB. Moreover, there is no quality improvement of the ``CsiNet+\_16Q4\_2/3\_64QAM'' when the channel SNR increases from $2$ dB to $10$ dB. Compared with these SSCC methods, the ADJSCC-CsiNet+ shows graceful performance improvement or degradation with the SNR increase or decrease. Even though the SSCC method works in the target SNR (e.g., the target SNR for the ``CsiNet+\_16Q4\_2/3\_64QAM'' is $2$ dB), the performance of the ADJSCC-CsiNet+ in the target SNR is at least 2 dB worse than the performance of the ``CsiNet+\_16Q4\_2/3\_64QAM''. Fig.~\ref{bw32_valid} with $k=32$ reveals similar results to Fig.~\ref{bw16_valid}. The gap between the performance of ADJSCC-CsiNet+ and the performance of the SSCC method with $k=32$ is larger than that with $k=16$.

\begin{figure}[!tb]
\centering
\includegraphics[width=0.9\columnwidth]{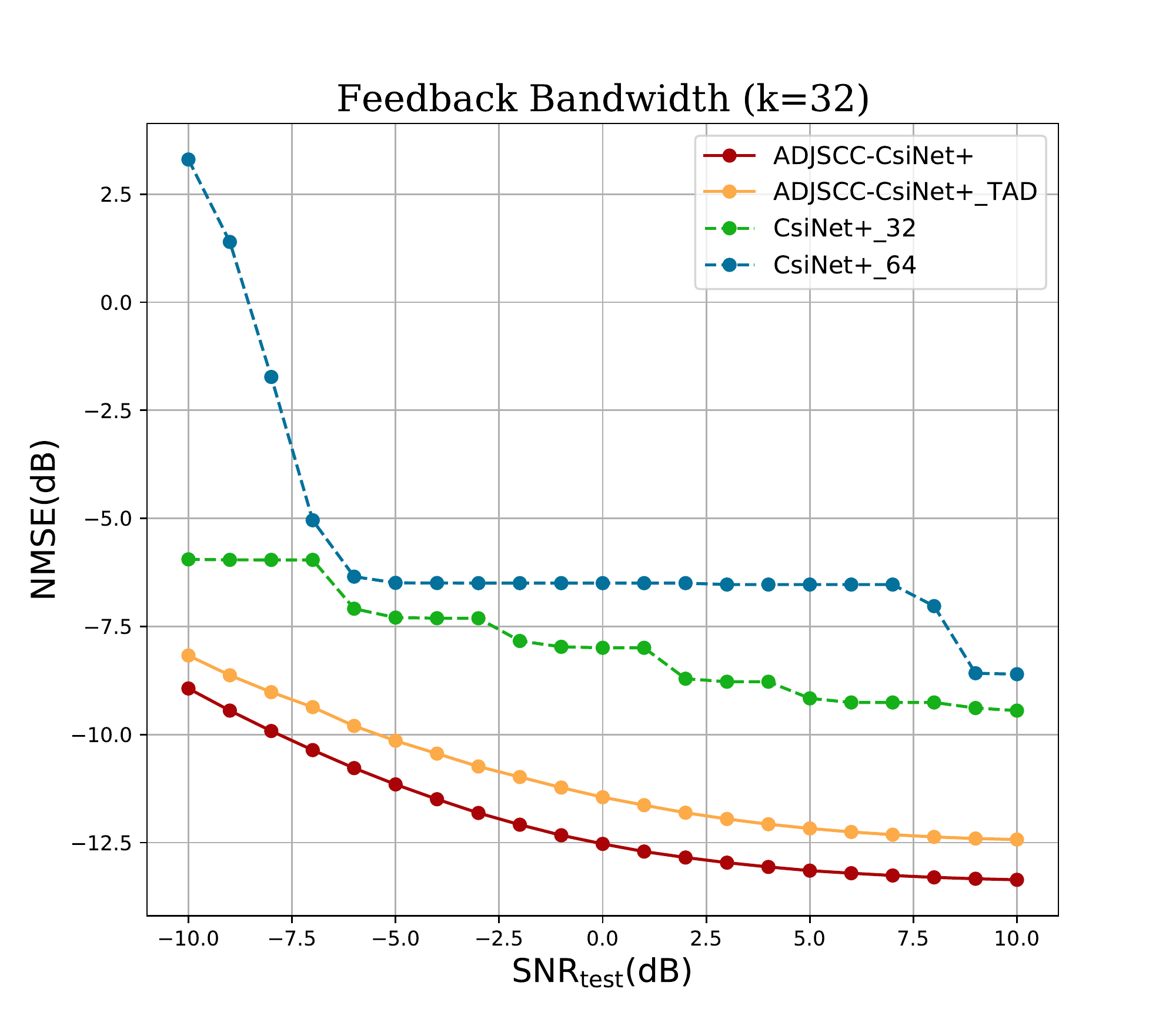}
\caption{The ablation study for ADJSCC-CsiNet+.
}\label{Fig:ablation}
\end{figure}

It worth noting that the CsiNet+ in SSCC methods is trained in the truncated angular-delay domain while the ADJSCC-CsiNet+ is trained in the spatial-frequency domain. Specifically, the ADJSCC-CsiNet+ uses the ATN to convert the CSI in the spatial-frequency domain to the transform domain with the same size as the truncated angular-delay domain in the UE. We infer that the performance gain of ADJSCC-CsiNet+ comes from both the non-linear transform and the DJSCC design philosophy. A thorough understanding of the ADJSCC-CsiNet+ is necessary to identify which part of the performance gain from the non-linear transform and  which part of the performance gain from the DJSCC design philosophy. Here, an ablation study is provided as shown in Fig.~\ref{Fig:ablation}. 

To focus on the performance gain, the performance of the SSCC methods with the same compressed dimensions of the CsiNet+ is simplified as an envelope curve, which consists of the best performance of these SSCC methods in the specific SNR. The curves ``CsiNet+\_32'' and ``CsiNet+\_64'' in Fig.~\ref{Fig:ablation} represent the envelope curves of the CsiNet+ encoders with 32 and 64 dimensional output, respectively. In other words, the curves of ``CsiNet+\_m'' can be regarded as the performance of the bit-level CsiNet+ combined with the AMC strategy. The curve ``ADJSCC-Csinet+\_TAD'' represents the proposed ADJSCC-Csinet+ without the ATN and the STN. The ``ADJSCC-Csinet+\_TAD'' is trained in the truncated angular-delay domain. As shown in Fig.~\ref{Fig:ablation}, the performance of the AJDSCC-CsiNet+\_TAD is $0.7\sim1.1$ dB worse than that of the the ADJSCC-CsiNet+ in $\rm SNR_{test} \in [-10,10]$ dB, which shows the performance gain of non-linear transform. The DJSCC performance gain is reflected by the performance gap between the ADJSCC-Csinet+ and the CsiNet+\_m. The performance of the AJDSCC-CsiNet+\_TAD is $2.2\sim3.6$ dB better than that of the CsiNet+\_32 and $3.4\sim11.4$ dB better than that of the CsiNet+\_64  in the $\rm SNR_{test}$ range from $-10$ dB to $10$ dB, respectively. Note that the CsiNet+ with $m=64$ uses more dimensions for compression than the CsiNet+ with $m=32$. The performance of the CsiNet+ with $m=64$ is definitely better than that of the CsiNet+ with $m=32$. Intuitively, the performance of the CsiNet+\_64 would be better than that of the CsiNet+\_32. However, the performance of the CsiNet+\_m depends on the matching strategies among the AMC, the length of quantized bits, and the output dimensions of the CsiNet+. If the length of quantized bits increases from $B$ to $B+1$, there are 32 additional bits for the quantized CsiNet+ with $m=32$ and 64 additional bits for the CsiNet+ with $m=64$. During the matching process between the quantized CsiNet+ and AMC, the quantized CsiNet+ with $m=32$ is more flexible for the AMC strategy than the quantized CsiNet+ with $m=64$. Therefore, as shown in Fig.~\ref{Fig:ablation}, the performance of the CsiNet+\_32 is better than that of the CsiNet+\_64.

\subsection{ADJSCC-CSI Adaptability Experiments}
\begin{figure}[!tb]
\centering
\includegraphics[width=0.9\columnwidth]{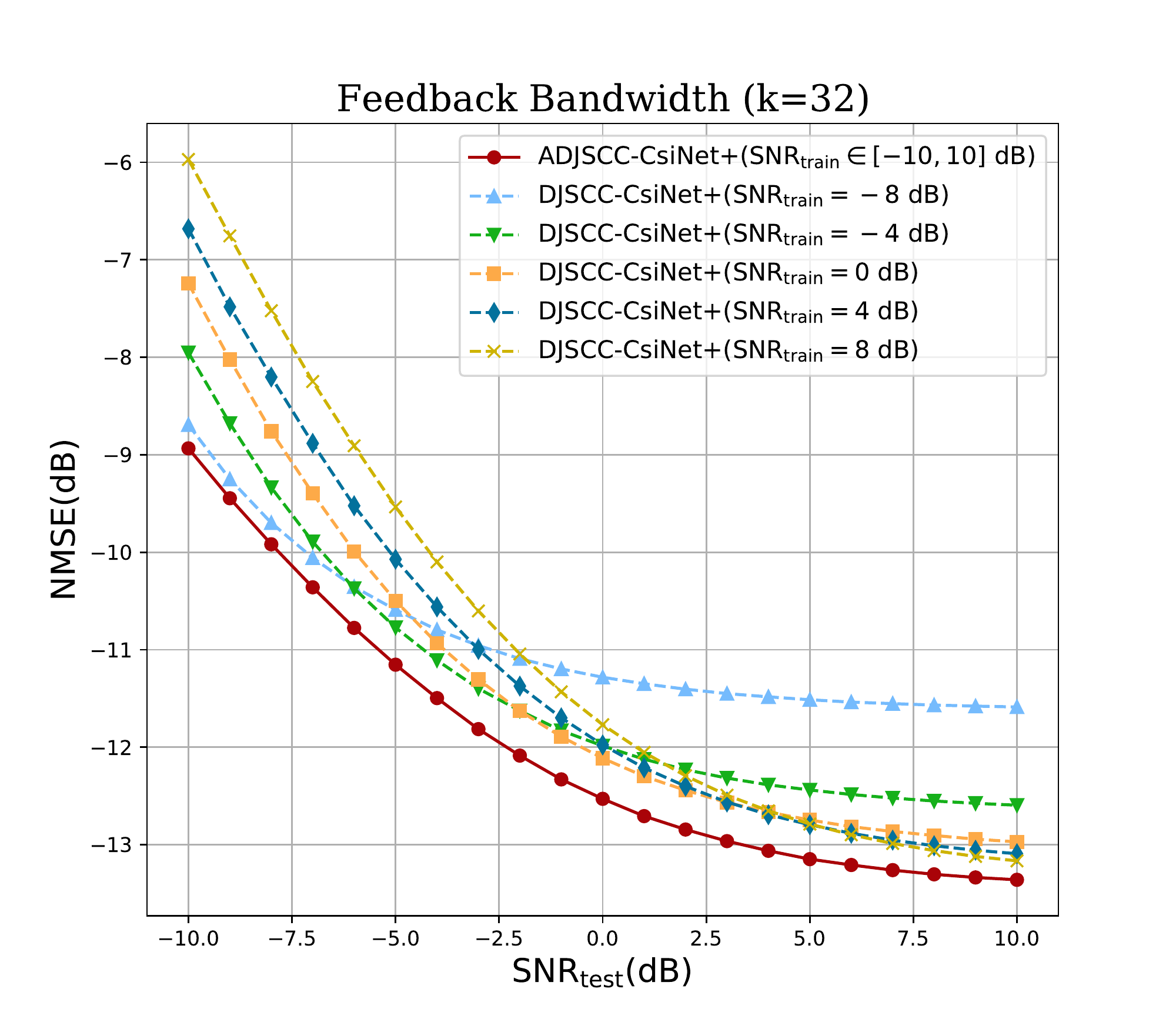}
\caption{Performance of ADJSCC-CsiNet+ and DJSCC-CsiNet+. The curve of ADJSCC-CsiNet+ is trained under the uniform distribution of SNR from -10dB to 10dB. Each curve of DJSCC-CsiNet+ is trained at a specific SNR.
}
\label{Fig:adapt}
\end{figure}

Now, we would like to demonstrate the effect of the AF Module only. Here, the DJSCC-CsiNet+ is introduced as a compared method. The DJSCC-CsiNet+ can be regarded as the ADJSCC-CsiNet+ without AF Modules. Fig.~\ref{Fig:adapt} compares the ADJSCC-CsiNet+ method with the DJSCC-CsiNet+ methods with feedback bandwidth $k=32$. The DJSCC-CSINet+ is trained at specific $\rm SNR_{train}=-8, -4, 0, 4, 8 $ dB, respectively. Both the ADJSCC-CsiNet+ and the DJSCC-CSINet+ are evaluated at specific $\rm SNR_{test} \in [-10,10]$ dB. Although the DJSCC can achieve remarkable performance when $\rm SNR_{test}$ is around $\rm SNR_{train}$, the performance of the DJSCC trained at specific $\rm SNR_{train}$ is not satisfactory when $\rm SNR_{test}$ is far from $\rm SNR_{train}$. For instance, the performance of the DJSCC-CSINet+ trained at $\rm SNR_{train}=-8$ dB has the best performance than other DJSCC-CSINet+ methods in $\rm SNR_{test}$ from $-10$ dB to $-7$ dB. With the increase of $\rm SNR_{test}$ from $-6$ dB to $-3$ dB, the performance of the DJSCC-CsiNet+ trained at $\rm SNR_{train}=-8$ dB is inferior to that of the DJSCC-CsiNet+ trained at $\rm SNR_{train}=-4$ dB. However, the performance of the ADJSCC-CsiNet+ outperforms the lower envelop of the provided DJSCC-CsiNet+ methods at any $\rm SNR_{test}$, which demonstrates the efficacy of AF Modules dealing with SNR adaption for the CSI feedback task.

\subsection{ADJSCC-CSI Generality Experiments}

Section \ref{subsection_4A} reveals that the CsiNet+ can be successfully converted to the DJSCC based method ADJSCC-CsiNet+ by employing the proposed ADJSCC-CSI framework. In this subsection, we apply the ADJSCC-CSI framework to the other two SC-CSI methods\textemdash the CsiNet proposed in \cite{wen2018deep} and the CRNet proposed in \cite{lu2020multi}\textemdash to show the generality of the proposed framework.
\begin{figure}[!tb]
\centering
\includegraphics[width=0.9\columnwidth]{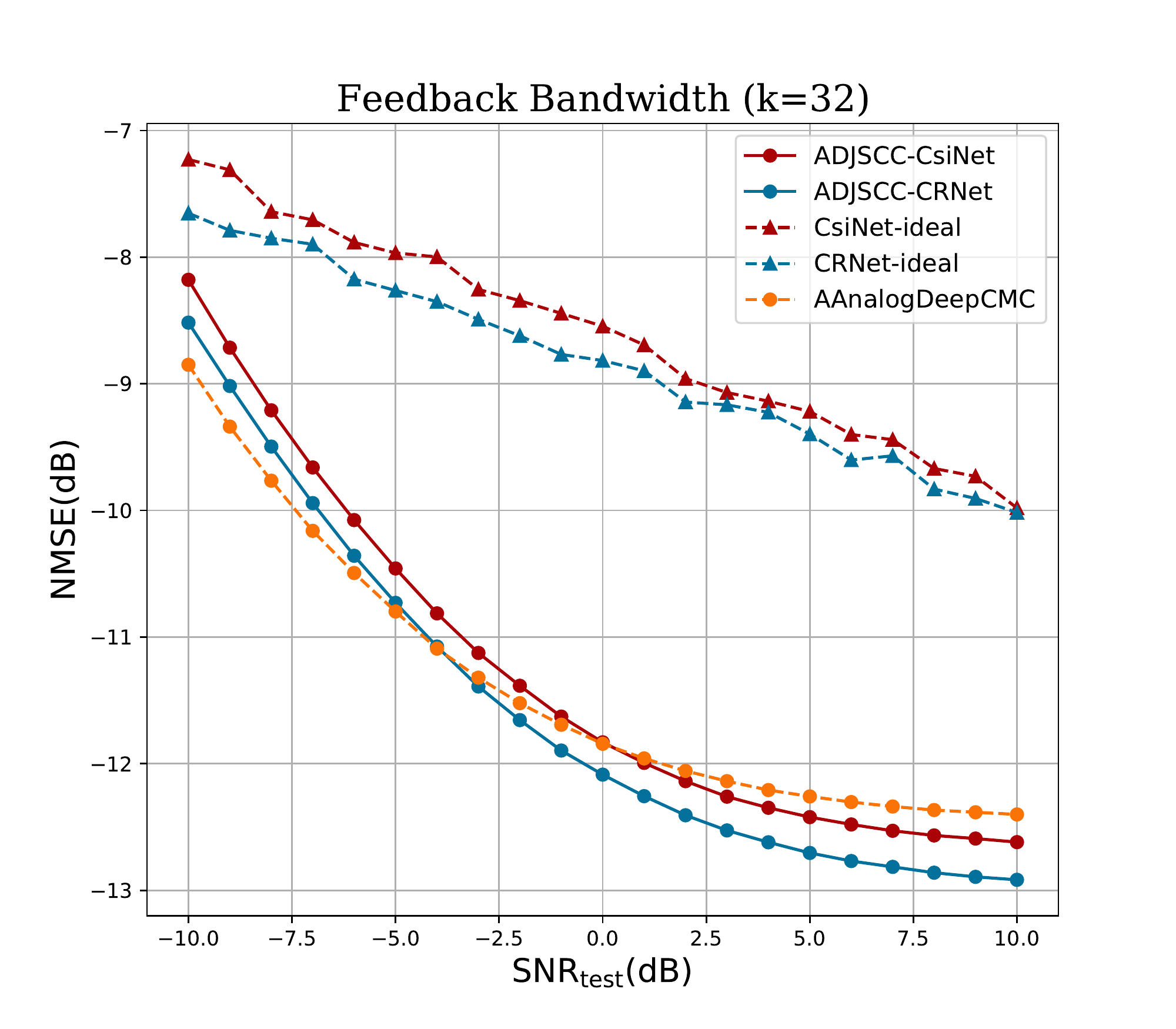}
\caption{Performance comparison of ADJSCC and ASSCC for CsiNet \cite{wen2018deep} and CRNet \cite{lu2020multi}. The AAnalogDeepCMC method proposed in \cite{mashhadi2020cnn} is plotted as reference.
}\label{Fig:generality}
\end{figure}
It is time consuming to traverse all of the possible combination schemes for the SSCC method to identify the best performance at the specific SNR. Here we employ some additional assumptions to reduce the difficulty in  evaluating the SSCC method. Our target is to compare the performance of the DJSCC based method with that of the SSCC based method, rather than to find the practical combination scheme for the SSCC method. Therefore, the channel capacity $C$\textemdash an upper bound for reliable transmission\textemdash instead of the AMC strategy is the employed  assumption to simplify the evaluation process for the SSCC method. Next, we consider about simplifying the quantization process. Due to no consideration of the quantization in \cite{wen2018deep} and \cite{lu2020multi}, the output of the SC-CSI-E at the UE is defaulted to 32-bit float values and then is assumed to be transmitted to the BS without error. The output bit length of the SC-CSI-E is $32\times m$ in \cite{wen2018deep} and \cite{lu2020multi}, where $m$ is the output dimension of the SSC-CSI-E. However, \cite{guo2020convolutional} demonstrates that the feedback of 32-bit float values is inefficient compared with the quantized version. According to the experiment results provided by \cite{guo2020convolutional}, the performance of the quantized version with $B=6$ is near the performance of the unquantized version represented by 32-bit float values. Here, we assume the performance of the quantized version with $B=5$ equals the performance of the unquantized version, which simplifies the quantization process and improves the upper bound of the performance of the SSCC method. Finally, the chosen dimension of the SC-encoder is expressed as:
\begin{equation} 
m=\lceil kC/B\rceil,
\end{equation}
where $k$ is the feedback bandwidth, $C$ is the feedback channel capacity for one channel use, $B$ is the assumed quantized bits, and $\lceil\cdot\rceil$ represents the ceiling operation. Then the performance of the SC-SCI method (e.g., the CSINet and the CRNet) with dimension $m$ is regarded as the performance of the SSCC method built upon the SC-SCI method with feedback bandwidth $k$. The ceiling operation further improves the upper bound of the performance of the SSCC method.

Fig. \ref{Fig:generality} compares the ADJSCC-CSI methods with the SSCC methods at feedback bandwidth $k=32$. Specifically, The labels ``CsiNet-ideal'' and ``CRNet-ideal'' represent the SSCC methods built upon the CsiNet and the CRNet under the three assumptions, respectively. The AnalogDeepCMC proposed by \cite{mashhadi2020cnn} also takes into account the uplink channel explicitly. Here, we insert AF Modules which we have introduced in Section \ref{subsection_3C} into the AnalogDeepCMC to construct the AAnalogDeepCMC for SNR adaption and plot its performance as reference. The hyper-parameter setting for the CsiNet and the CRNet is the same as that for the ADJSCC-CsiNet+. Based on the aforementioned assumptions, the performance of the CsiNet-ideal and the CRNet-ideal are the upper bound of the SSCC based Csinet and the SSCC based CRNet, respectively. Nonetheless, the performance of the CsiNet-ideal and the CRNet-ideal are still worse than the performance of the ADJSCC-CsiNet and the ADJSCC-CRNet, respectively. Specifically, the performance of the CsiNet is 0.9 dB worse than the performance of the ADJSCC-CsiNet at $\rm SNR_{test}=-10$ dB. With the increase of the $\rm SNR_{test}$ from $-10$ dB to $1$ dB, the ADJSCC-CsiNet brings a gradually increased performance, outperforming the CsiNet-ideal by a margin at almost $3.3$ dB. With the further increase of the $\rm SNR_{test}$ from $1$ dB to $10$ dB, the gap between the CsiNet-ideal and the ADJSCC-CsiNet slowly decreases, while is still larger than $2.6$ dB. In other words, if $\rm NMSE=-10$ dB is required at the BS to promise the reconstruction quality, the UE adopting the ADJSCC-CsiNet can save at least 14 dB transmission power than that adopting the CsiNet-ideal. The comparison between the CRNet-ideal and the ADJSCC-CRNet reveals similar results.

It has been demonstrated that the performance of the CRNet is better than that of the CsiNet in \cite{lu2020multi}. Hence the performance of the CRNet-ideal should be better than that of the CSiNet-ideal. This is consistent with the comparison result between the CRNet-ideal and the CSiNet-ideal revealed in Fig. \ref{Fig:generality}. Moreover, the comparison between the ADJSCC-CRNet and the ADJSCC-CsiNet in Fig. \ref{Fig:generality} reveals a similar trend. We infer that is because the ability of CRNet is stronger than that of CsiNet. Compared with the AAnalogDeepCMC, the ADJSCC-CsiNet is inferior in the $\rm SNR_{test}$ range from $-10$ dB to $0$ dB and is superior in in the $\rm SNR_{test}$ range from $1$ dB to $10$ dB. The comparison between the AAnalogDeepCMC and the ADJSCC-CRNet shows similar results except the partition point changed from $0$ dB to $-4$ dB.

\begin{table}[!tb] 
\renewcommand{\arraystretch}{1.3} 
\caption{storage comparison for the CSI feedback task} 
\label{Table:storage}
\centering
\begin{tabular}{c|c|c}
\hline
\textbf{Method} & \textbf{UE params}& \textbf{BS params}\\ \hline
ADJSCC-CsiNet & \textbf{233,874} & \textbf{241,280} \\ \hline
ADJSCC-CRNet & 234,260 & 241,552 \\ \hline
CsiNet-ideal & 999,300 & 1,122,882 \\ \hline
CRNet-ideal & 999,846 & 1,103,562  \\ \hline
AAnalogDeepCMC & 1,791,656 & 8,416,962 \\ \hline
\end{tabular}
\end{table}

We finally calculate the parameters needed in the UE and in the BS for the aforementioned methods, respectively. In the practical wireless scenario, many CsiNet models should be prepared for the CsiNet-ideal to cover the SNR range from $-10$ dB to $10$ dB. However, one model is enough for the ADJSCC-CSI method and the AAnalogDeepCMC method due to the SNR adaption mechanism. Assuming one model corresponds to a specific SNR for the CsiNet-ideal to keep the optimality. The parameter sizes of these methods are given in Table \ref{Table:storage}. As we can see, the parameter size of the CsiNet-ideal is over 4$\times$ heavier than that of the ADJSCC-CsiNet both in the UE and in the BS. The parameter size of the AAnalogDeepCMC is over 7$\times$ and 34$\times$ heavier than that of the ADJSCC-CsiNet in the UE and in the BS, respectively. The parameter sizes of the CRNet-ideal and the ADJSCC-CRNet are slightly more than that of the CsiNet-ideal and the ADJSCC-CsiNet, respectively. To summarize, the proposed ADJSCC-CSI is storage friendly and achieves better performance than the SSCC based method for the CSI feedback task.


\section{Conclusion}

In this work, we propose a novel ADJSCC-CSI framework for the CSI feedback task, which includes three key components, e.g., non-linear transform, DJSCC and SNR adaption. The proposed framework can successfully upgrade the existing SC-CSI method (e.g., CsiNet, CsiNet+, CRNet) to a DJSCC based method. During the end-to-end training, ADJSCC-CSI framework can learn a pair of transformation networks, a pair of encoder and decoder built upon the SC-CSI method, and SNR adaption strategy. 

In the experiments, we compared the ADJSCC-CsiNet+ with SSCC based methods and the DJSCC based methods built upon CsiNet+ to show its validity and adaptability, respectively. Then we apply the proposed ADJSCC-CSI framework to the CsiNet and the CRNet to demonstrate its generality. In addition, compared with the SSCC based method, the ADJSCC-CSI framework can save more storage both in the UE and in the BS. 

In future work, a potential direction is to jointly design the channel estimation module and the CSI feedback module with the DJSCC method, which might further reduce the CSI feedback bandwidth on the basis of acceptable CSI reconstruction performance.

\bibliographystyle{IEEEtran}
\bibliography{ref/ref.bib}

\end{document}